\documentclass{article}

\usepackage{PRIMEarxiv}

\usepackage[utf8]{inputenc} 
\usepackage[T1]{fontenc}    
\usepackage{hyperref}       
\usepackage{url}            
\usepackage{booktabs}       
\usepackage{amsfonts}       
\usepackage{nicefrac}       
\usepackage{microtype}      
\usepackage{lipsum}
\usepackage{fancyhdr}       
\usepackage{graphicx}       
\graphicspath{{media/}}     

\usepackage{subfig}
\usepackage{amsmath}
\usepackage{siunitx}

\title{Magnetorheological Axisymmetric Actuator with Permanent Magnet
}

\author{
  Jakub Bernat, Paulina Superczynska \\
  Institute of Automatic Control and Robotics \\
  Poznan University of Technology \\
  Poznan, Poland\\
  \texttt{\{jakub.bernat, paulina.superczynska\}@put.poznan.pl} \\
   \And
  Piotr Gajewski, Agnieszka Marcinkowska \\
  Institute of Chemical Technology and Engineering \\
  Poznan University of Technology \\
  Poznan, Poland\\
  \texttt{\{piotr.gajewski, agnieszka.marcinkowska\}@put.poznan.pl} \\
}

\begin{document}
\maketitle

\begin{abstract}
This study examines the concept of axisymmetric actuator based on the magnetorheological membrane, electromagnet and permanent magnet. The construction of the actuator enables its application in wide range of practical devices like pumping, loudspeaker or varying-stiffness button. This work will highlight its working principle especially the influence of permanent magnet. Furthermore, the model of devices will be defined relaying on the Hammerstein model. To show the properties of the actuator and to perform the model identification, the set of experiments was run taking into account static and dynamic working conditions.
\end{abstract}


\section{Introduction}

In recent years, actuators made of intelligent materials have gained considerable interest among automation specialists \cite{alkhalaf2020composite,kashima2012novel}.  Actuators for which the stimulus causing the action (deformation or displacement) is heat, magnetic or electric field are made of soft and flexible intelligent materials susceptible to this type of force. Examples of such devices are magnetorheological elastomers (MREs) for which a magnetic field generates a reaction. Due to their structure (ferromagnetic particles embedded in polymer elastomers) these actuators have both ferromagnetic and viscoelastic properties \cite{kashima2012novel}. Recently, these kinds of materials are widely used in many applications. For example, in biological soft robotic and magneto-active substrates for non-invasive cell stimulation or as drug delivery systems \cite{moreno2022magneto,kim2019ferromagnetic,koivikko2021magnetically,yarali2022magneto,uslu2021engineered}, in damping systems such as multifunctional grille composite sandwich plates which can be used in aerospace or marine area \cite{li2021vibration} or as base isolator in buildings and bridges to resist unpredictable loading conditions e.g. seismic activities \cite{khayam2020development}. Moreover, they can be used as smart sensors \cite{alkhalaf2020composite} and microfluidic devices \cite{amiri2021vibration,moreno2022hybrid}.

MREs are materials consisting of polymeric elastomers that provide viscoelastic properties, and ferromagnetic particles that are responsible for magnetic properties \cite{polym12123023}. By choosing the proper polymer matrix the MRE could be highly elastic materials that are able to interact in a magnetic field. The typical elastomeric matrix could be silicone rubber \cite{BICA20121666,SHENOY2020166169}, polyurethane rubber \cite{Wei2010,ma13204597}, nitrile rubber \cite{LOKANDER2003245}, polybutadiene rubber \cite{SUN2008520}, and others (odnośnik). Depending on the type of used elastomer, the polymer cross-linking density, the presence of other additives and etc. it is possible to obtain materials with a wide range of properties (Young modulus, hardness, elasticity, etc.) \cite{Zhao_2019}. By introducing ferromagnetic particles in various concentrations, sizes etc. it is possible to affect the final magnetic properties of the MRE material \cite{BASTOLA2020108348}. That gives the possibility of significant modification of the target material properties depending on the requirements.

The modeling of soft devices can be grouped into geometry design models and control-oriented models. The purpose of the first one is to model the physical phenomena in the device to exploit its working principle or to optimize the geometry. The Finite Element Method (FEM) models of the MRE matarial requires the modeling of electromagnetic properties and mechanical properties \cite{Cantera_2017,Yarali2020}. The goal of the second group of models is to find the properties of the device required to design the controller. In the case of MRE this subject is less exploited. Most models are related to vibration control as described in \cite{NGUYEN2018192,Li2010733,Shou2022,Zapomel2017191,Nguyen2022}. Nonetheless, the control-oriented models of different smart actuators exist in the literature. For instance, the DEAP actuator is presented in works \cite{Rizzello:6867294,Bernat:9293025} where the identification and modeling are performed. The common factor of DEAP and MRE are similar types of membranes - in both cases, the sillicone-based membrane is possible to use. 
In the case of the DEAP actuator, the popular identification procedure is performed by multiple optimization processes considering different physical phenomena  \cite{Rizzello:6867294,Bernat:9293025}. Alternative work in this field is presented in \cite{electronics10111326} where the model of the DEAP actuator is split into fast and slow dynamics. Therefore, the identification can be performed in two steps. In our work, the identification method is a generalization of the procedure presented in \cite{electronics10111326}.

In general, the identification of dynamic systems is a well-defined part of control system theory \cite{soderstrom1989system,ljung1999system,Wills201370}. There exists a lot of models like transfer functions \cite{ljung1999system}, Hammerstein-Wiener models \cite{mzyk2013combined,Wills201370} and NARMA models \cite{Narendra1997475} which enables to describe the dynamics of the systems. The main properties of such models are linearity/nonlinearity, time variance or invariance, discrete/continuous, and the existence of disturbance or noise. The general identification algorithms exist for mentioned general models like Recursive-Least-Square method, Model Prediction Error, or Instrumental Variable. There are also optimization-based methods like the Maximum Likehood method \cite{soderstrom1989system,ljung1999system}. In our work, we focused on identifying a particular type of dynamic system, an example of which is the proposed MRE actuator, where standard identification methods are not well suited. Therefore, the new identification algorithm exploiting the features of MRE actuator is defined.

The main goal of this work is to show the properties of the MRE actuator. Its geometry is axisymmetrical which is inspired by DEAP actuators \cite{Rizzello:6867294}, which should enable a wide range of motion of the prepared MRE through the additional use of permanent magnets. To explore the properties of the actuator, the series of experiments is conducted. Furthermore, the control oriented model is proposed and identified based on the experimental data. The actuator model is based on the Hammerstein model which is well known in the identification theory. 

Besides of Introduction, the work consists of the following sections:  Section \ref{sec:concept}, in which the concept of actuator is shown, Section \ref{se:hammersteinModelAndIdentification} describing its model and identification algorithm and finally, the experiments and conclusions in the Section \ref{se:experiments} and \ref{se:conclusion}, respectively.

\section{Concept of Actuator}
\label{sec:concept}

\begin{figure}
     \centering
     \subfloat[][]{\includegraphics[width=0.45\textwidth]{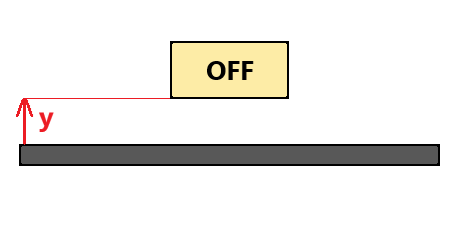}\label{fig:actuator:a}}
     \subfloat[][]{\includegraphics[width=0.45\textwidth]{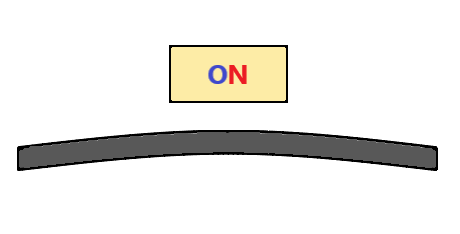}\label{fig:actuator:b}}
     \\
     \subfloat[][]{\includegraphics[width=0.45\textwidth]{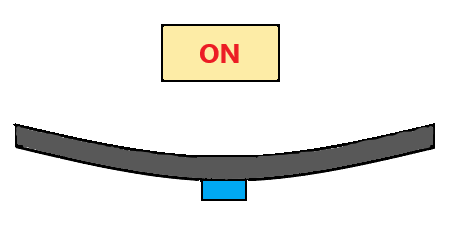}\label{fig:actuator:c}}
     \subfloat[][]{\includegraphics[width=0.45\textwidth]{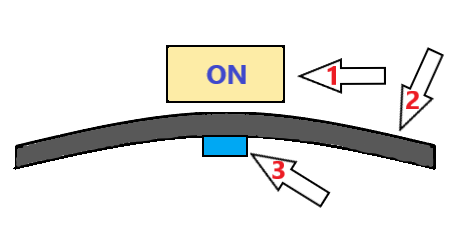}\label{fig:actuator:d}}
     \caption{The schema of the actuator: a) electromagnet off, b) electromagnet on, $I<0$, c) electromagnet on, I>0,permanent magnet added, d) electromagnet on, $I<0$, permanent magnet added. Arrows: 1- electromagnet, 2- membrane, 3- permanent magnet.}
     \label{fig:actuator}
\end{figure}

In this section, the idea of the MRE actuator is presented. The main features of that kind of device are softness and reaction on the magnetic field. The actuator consists of two main components. Membrane - a moving part, made of the material sensitive to the magnetic field and an electromagnet that forces the object to move. The membrane is placed between two specially cut pieces of plexiglass that keep it in the right position and prevent it from moving. By applying power to the electromagnet, the membrane is forced to move toward the electromagnet, causing it to deflect. In order to achieve a greater operating range of the actuator, it is possible to add an element that increases the electromagnetic interaction - a permanent magnet. Figure \ref{fig:actuator} presents all 3 mentioned parts of the actuator where arrows points on: 1- electromagnet, 2- membrane, 3- permanent magnet. Figure \ref{fig:actuator}a) presents the situation when the electromagnet is off, and there are no changes in the system. Figure \ref{fig:actuator}b) presents scenario when electromagnet is on and $I<0$ (blue) or red ($I>0$). Figure \ref{fig:actuator}c) and d) presents the same scenario but with two possible modes of electromagnet operations: red represents the situation when $I>0$, and blue when $I<0$. In both situations a permanent magnet is applied and the deflection is more noticeable due to the greater interaction of field forces and depending on operation mode membrane moves towards or away from the electromagnet.

From the control point of view, the device has single input which is voltage $u$ applied to the electromagnet. The output $y$ is the distance between the electromagnet and MRE membrane. It can be treated as a global coordinate. Additionally, the auxiliary local coordinate $\Delta y = y - y_0$ is introduced where $y_0$ is position of membrane for zero voltage $(u=0)$. In our work, the position $y_0$ depends on the number of permanent magnets applied to the device.

\section{Hammerstein Model and Its Identification}
\label{se:hammersteinModelAndIdentification}

\begin{figure}
     \centering
     \subfloat[][local coordinates]{\includegraphics[width=0.45\textwidth]{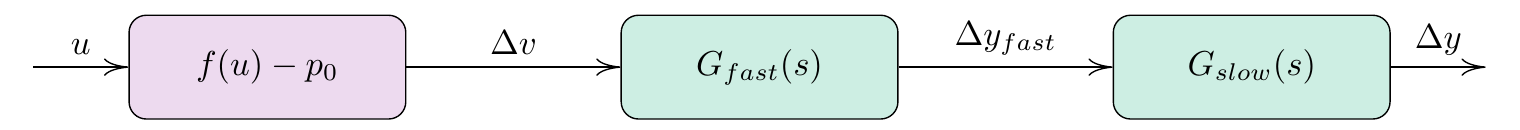}\label{fig:schema-local}}
     \\
     \subfloat[][global coordinates]{\includegraphics[width=0.45\textwidth]{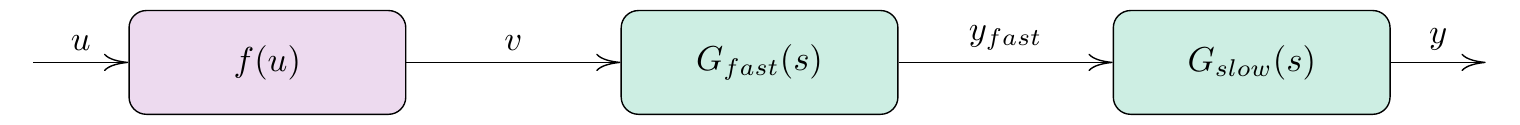}\label{fig:schema-global}}
     \caption{The schema of the Hammerstein nonlinear system with the sliced transfer functions on slow and fast.}
     \label{fig:schema}
\end{figure}

In this section, the Hammerstein model of MRE axisymmetrical actuator is presented with its identification procedure. The model of the device consists of static non-linearity and transfer function describing dynamics. We will show that the MRE actuator allows to divide transfer function into slow and fast dynamics. The schema of the model is illustrated in Figure \ref{fig:schema} where the definition of blocks presented in schema is as follows.
The nonlinear $f(u)$ is expressed as $p$ order polynomial:
\begin{equation} \label{eq:staticNonlinearity}
    f(u) = \sum_{k=0}^{p}{p_{k} u^k}.
\end{equation}
where $p_k$ are the polynomial coefficients and $p$ is the polynomial order. The parameter $p_0$ in polynomial can be applied to transfer from global coordinate $y$ to local coordinate $\Delta y$. If we set voltage to zero ($u=0$) in the static nonlinearity, then it gives that $y_0 = f(0) = p_0$, and hence $\Delta y = y - p_0$.

The dynamics are described by two transfer functions. The fast dynamics are given by:
\begin{equation}
G_{fast}(s) = \frac{k\left( \omega^2 + \alpha^2 \right)}{s^2 + 2\alpha s + \alpha^2 + \omega^2}
\end{equation}
where $k$ is the gain, $\omega$ is the damped natural frequency, $\alpha$ is the decay ratio. Relying on the experimental knowledge, the fast dynamics is specified as the under-dumped system. The slow dynamics are equal to:
\begin{equation}
G_{slow}(s) = \frac{s + z_0}{s + s_0}
\end{equation}
where $z_0$ is the zero and $s_0$ is the pole. The static gain of the series connection of both systems is assumed to be one:
\begin{equation} \label{eq:staticGain}
G(0) = G_{fast}(0)G_{slow}(0) = 1.
\end{equation}
Therefore, the parameters of both transfer functions are connected with the formula:
\begin{equation} \label{eq:parameterConstraint}
   k \frac{z_0}{s_0} = 1.    
\end{equation}

\subsection{Identification}

In the identification process, there are made preliminary assumptions based on the observations in the experiments. The dynamics of slow is not important in the range of high frequencies which means that for $f >> 1$, transfer function $G_{slow}(s) \approx 1$. On the other hand, $f << 1$, transfer function $G_{fast}(s) \approx k$.

The data required in the identification process are as follows:
\begin{itemize}
   \item the static characteristic between input and output expressed as set of points $U_i$, $Y_i$, where $i=1,\ldots,P_{s}$ 
   \item the set of step responses measured for input change from voltage $u^j_{pre}$ to $u^j_{post}$ and output $\Delta y^j_{step}(t_i) = y(t_i) - y(0) = \Delta y(t_i) - \Delta y(0)$, where $j=1,\ldots,P_{d}$. The output is only known for fixed time points $t_i$, where $i=1,\ldots,N$
\end{itemize}
The experimental procedure for acquiring data will be shown in the following section. The required data can be easily measured in the experiment.

In the first step, the static nonlinearity $f(u)$ is analyzed. Thanks to the static gain of the systems $G_{fast}(0)G_{slow}(0)$ is equal to 1, it is possible to identify only the nonlinear term. In our work, to find the coefficients $p_{k}$ the curve fitting problem is solved by comparing the measurements $U_i$, $Y_i$ for the steady state: 
\begin{equation} \label{eq:problemStaticNonlinearity}
    \min_{p_0, \ldots, p_k}{\sum_{i=0}^{P}{\left[ f(U_i) - Y_i \right]^2 }}
\end{equation}
The degree of polynomial $p$ can be chosen, for instance, by solving the above optimization problem multiple times with different $p$ values and choosing the best one. In general, the function $f(u)$ is different for various configurations of the device like varying permanent magnet and varying MRE material. Knowing the nonlinear term $f(u)$ defined in (\ref{eq:staticNonlinearity}), it is possible to compensate it and hence directly identify the dynamic model $G_{fast}(s)$ and $G_{slow}(s)$.

In the second step, parameters of the transfer function $G_{fast}(s)$ are found. The input to transfer function is signal $v$ which can be calculated only on the basis of input $u$. Furthermore, it is possible to calculate the formula for the step response of the fast system $G_{fast}(s)$: 
\begin{equation}
\eta_{fast}(t,k,\alpha,\omega) = k \left[ 1 + \frac{\sqrt{\alpha^2 + \omega^2}}{\omega}\cos\left( \omega t + \varphi_z \right) e^{-\alpha t}  \right]
\end{equation}
where $\varphi_z = \textrm{atan2}(-\omega, -\alpha) - \frac{\pi}{2}$. Therefore, the identification can be solved by optimization problem:
\begin{equation} \label{eq:problemGfast}
    \min_{k^j,\alpha^j,\omega^j} \frac{1}{N_{fast}}{\sum_{i=0}^{N_{fast}}{\left| \eta_{fast}(t_i,k^j,\alpha^j,\omega^j) \Delta v^j_{step} - \Delta y^j_{step}(t_i) \right|}}
\end{equation}
where $\Delta v^j_{step} = f(u^j_{\text{post}}) - f(u^j_{\text{pre}})$. $N_{fast} << N$ is the number of points in the fast response.

The optimization problem is solved for multiple-step responses independently. Therefore, for each response denoted as $j$, we obtain the parameters $k^j$, $\alpha^j$ and $\omega^j$. Hence, we obtained a list of parameters from which we have to calculate the estimates of model parameters. In our work, we have determined the final parameters by calculating the median value from all parameters:
\begin{equation}
\begin{aligned}
    \tilde{k} &= \text{med}\left(\left\{ k^j, j=1,\ldots,P_d \right\}\right) \\
    \tilde{\alpha} &= \text{med}\left(\left\{ \alpha^j, j=1,\ldots,P_d\right\}\right) \\
    \tilde{\omega} &= \text{med}\left(\left\{ \omega^j, j=1,\ldots,P_d \right\}\right)
\end{aligned}
\end{equation}
We chose the median because it is less sensitive for single high distortions in comparison to the average. In our simulation practice, the distortions were the results of optimization which can give sometimes the local optimum to be far away from true values. 

In the last step, the parameters of transfer function $G_{slow}(s)$ are found. It is required to find only single parameters $s_0$ or $z_0$ because of $k \cdot \frac{z_0}{s_0} = 1$ from (\ref{eq:staticGain}). The unit step response for $G_{slow}(s)$ can be expressed as:
\begin{equation} 
\eta_{slow}(t,s_0) = \frac{z_0}{s_0} - \frac{z_0 - s_0}{s_0}e^{-s_0 t}
\end{equation}
\begin{equation} \label{eq:problemGslow}
    \min_{s^j_0} \frac{1}{N}{\sum_{i=0}^{N}{\left| \eta_{slow}(t_i,s^j_0) k \Delta v_j - \Delta y^j_{step}(t_i) \right|}}
\end{equation}
The term $\Delta v_j$ is multiplied by $k$ because of the influence of steady-state gain of $G_{fast}(0)$. In this case, similarly to the searching $\tilde{k}$, $\tilde{\alpha}$ and $\tilde{\omega}$, we have also run the optimization problem for multiple step responses. Therefore, the median is also taken for $s^j_0$ as:
\begin{equation}
 \tilde{s}_0 = \text{med}\left(\left\{ s^j_0, j=1,\ldots,P_d \right\}\right)
\end{equation}
The last parameter $\tilde{z}_0$ is known from the relation (\ref{eq:parameterConstraint}) and hence $\tilde{z}_0 = \frac{\tilde{s}_0}{\tilde{k}}$. 

In summary, the identification procedure has three main steps. The first is searching the shape of static nonlinearity by solving the equation (\ref{eq:problemStaticNonlinearity}). The second is to find the coefficients of $G_{fast}(s)$ in the problem (\ref{eq:problemGfast}). And the last step is to get $G_{slow}(s)$ by solving problem (\ref{eq:problemGslow}).

\section{Experiments}
\label{se:experiments}

This section describes the experiments performed on the proposed actuator. Firstly, the construction of the actuator is explained. Next, the experimental tests of the device and the identification results are presented. 

\subsection{Construction}

\begin{figure}
  \centering
  \includegraphics[width=0.6\textwidth]{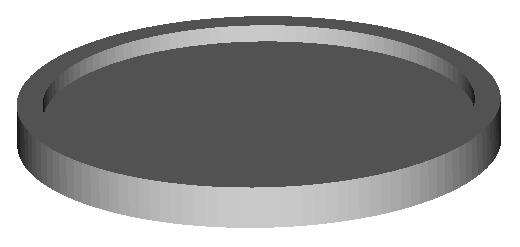}
  \caption{The mold for magnetorheological membrane.}
  \label{fig:molds}
\end{figure}

The magnetorheological membranes were prepared according to the previously described procedure \cite{self:gripperMRE2022}. The RTV-2 silicone (OTT-S825 from OTTSilicone) was mixed with iron powder in mass proportion 3:7. The silicone and iron were thoroughly mixed to obtain a homogeneous mixture. Then the \SI{2}{\percent} by weight of OTT-S825 catalyst in relation to the amount of silicone was introduced. After mixing, the prepared mixture was degassed by vacuum treatment to remove any air bubbles. This allowed for avoiding the bubbles in the MRE sample. Then, the mixture was poured into the prepared molds (Figure \ref{fig:molds}) and left for the next 12 h to cross-linking.

\begin{figure}
  \centering
  \includegraphics[width=0.9\textwidth]{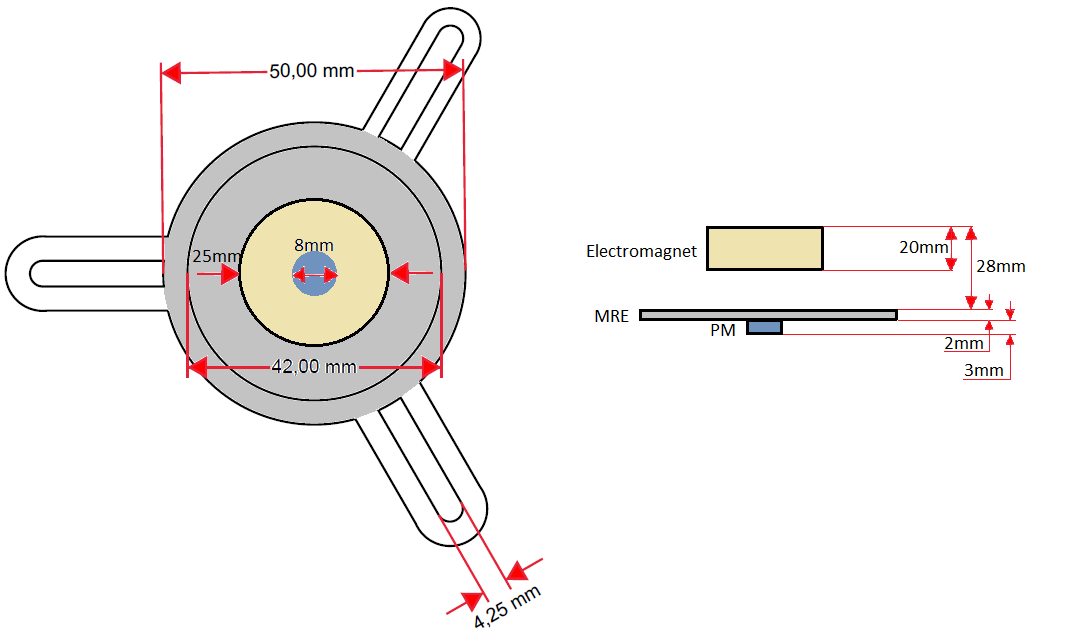}
  \caption{The dimensions of magnetorheological axisymmetric actuator.}
  \label{fig:laboratory_kit_3}
\end{figure}

The design and all dimensions of the laboratory kit are presented in Figure \ref{fig:laboratory_kit_3}. Parameters of the used components are collected in a Table \ref{tab:parameters}.

 \begin{table*}[htbp]
 \renewcommand{\arraystretch}{1.6} 
 \caption{Parameters of the used components.}
 \begin{center}
\begin{tabular}{c| c| c| c}
\hline
\multicolumn{2}{c|}{\textbf{Electromagnet}}
& \multicolumn{2}{c}{\textbf{Permanent Magnet}}\\
\cline{1-4} 
\textbf{\textit{Parameter}}
& \textbf{\textit{Value}}
& \textbf{\textit{Parameter}}
& \textbf{\textit{Value}} \\ 
\cline{1-4} 
\hline
Supply voltage & 12V & Magnetic induction &  0,304T\\
\hline
Operational current & 140mA & Magnetic moment & 171,684nWb x m\\
\hline
Lifting force & 8kg & Maximum lifting capacity & 1,2kg\\
\hline
Diameter & 25mm & Diameter & 8mm\\
\hline
Height & 20mm & Height & 3mm\\
\hline
\end{tabular}
\label{tab:parameters}
\end{center}
\end{table*}

\begin{figure}
  \centering
  \includegraphics[width=0.6\textwidth]{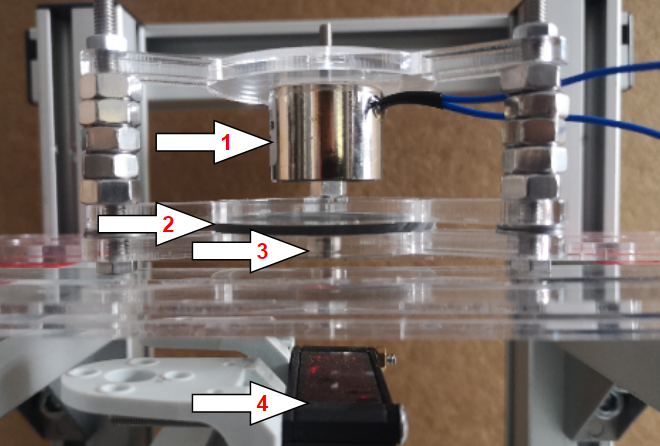}
  \caption{Laboratory kit with magnetorheological axisymmetric actuator. Arrows: 1- electromagnet, 2- membrane, 3- permanent magnet, 4- laser.}
  \label{fig:laboratory_kit_1}
\end{figure}

Figure \ref{fig:laboratory_kit_1} presents a laboratory kit with an electromagnet that is used to collect the results. The important elements are marked: 1 - electromagnet, 2 - MRE, 3 - permanent magnet, and 4 - laser.
As mentioned previously the main part of this actuator is a membrane which is its movable part. And as it is described, it is made of a material sensitive to the magnetic field. The membrane placed between two pieces of plexiglass starts to deflect after applying the power to the electromagnet. During experiments, permanent magnets were added underneath the membrane to achieve a greater operating range for the actuator.

\begin{figure}
  \centering
  \includegraphics[width=0.6\textwidth]{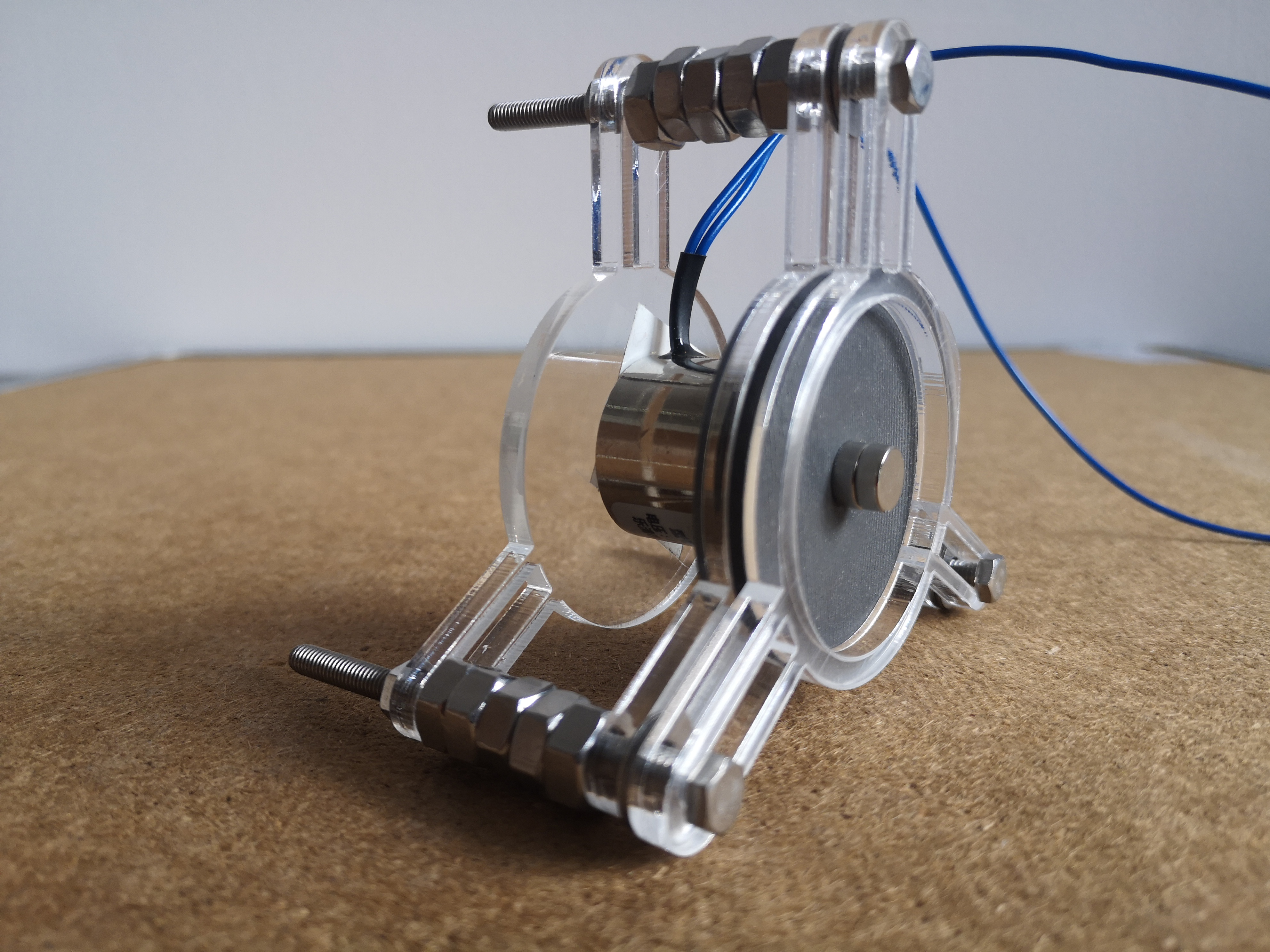}
  \caption{The standalone magnetorheological axisymmetric actuator.}
  \label{fig:laboratory_kit_2}
\end{figure}

Figure \ref{fig:laboratory_kit_2} presents the main part of the laboratory kit- MRE axisymmetric actuator placed between two pieces of plexiglass with a permanent magnet placed in its center, and electromagnet on the opposite side.

\subsection{Experiments}

\begin{figure}
  \centering
  \includegraphics[width=0.6\textwidth]{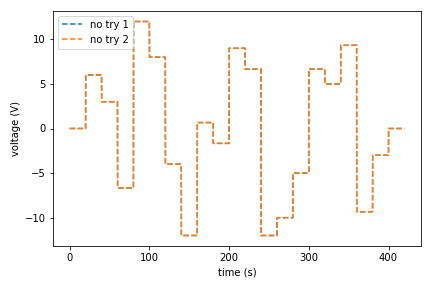}
  \caption{The voltage excitation applied to the actuator to measure step responses.}
  \label{fig:voltage_steps}
\end{figure}

\begin{figure}
     \centering
     \subfloat[][single magnet - all]{\includegraphics[width=0.45\textwidth]{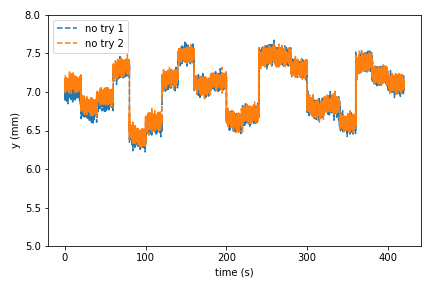}\label{fig:steps1}}
     \subfloat[][double magnet - all]{\includegraphics[width=0.45\textwidth]{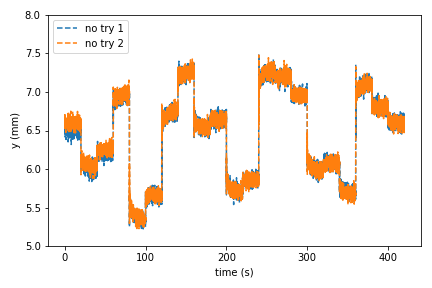}\label{fig:steps2}}
     \\
     \subfloat[][single magnet - zoom long]{\includegraphics[width=0.45\textwidth]{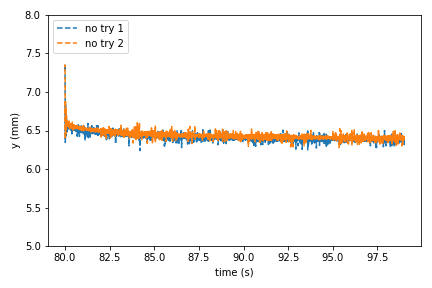}\label{fig:steps1_zoom_long}}
     \subfloat[][double magnet - zoom long]{\includegraphics[width=0.45\textwidth]{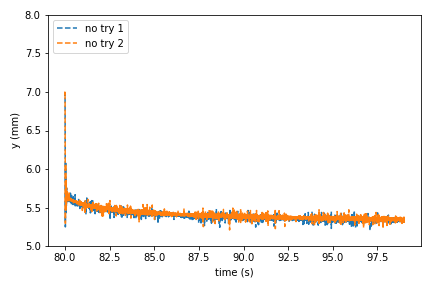}\label{fig:steps2_zoom_long}}
     \\
     \subfloat[][single magnet - zoom short]{\includegraphics[width=0.45\textwidth]{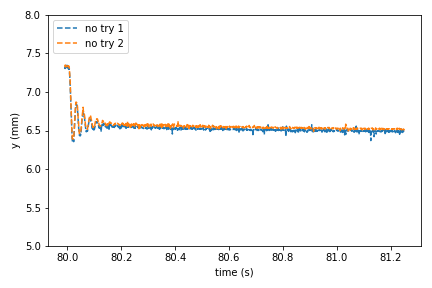}\label{fig:steps1_zoom}}
     \subfloat[][double magnet - zoom short]{\includegraphics[width=0.45\textwidth]{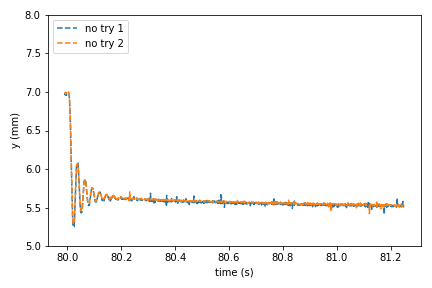}\label{fig:steps2_zoom}}
     \caption{The step responses for actuators with various number of magnets.}
     \label{fig:steps}
\end{figure}

This section examines the performed experiments to show the properties of the MRE axisymmetric actuator. The characteristic was measured by performing a set of step responses of the actuator as it is shown in Figure \ref{fig:voltage_steps}. The period of switches is long enough to obtain the steady state by the actuator. The range of the voltage is from \SIrange{-12}{12}{\volt}, the period is equal to \SI{20}{\second} and the total time of input was \SI{420}{\second}. The response was measured for none, one, and two permanent magnets. All characteristics were measured two times to check the repeatability. The overall response is presented in Figure \ref{fig:steps} a-b for one and two permanent magnets (the characteristics of the actuator without magnets are almost constant, and are not shown here, however they will be analyzed in the static section). To allow further processing, the response was split into separate step responses. The total number of step responses was \si{20}. The example of the single response is shown in Figure \ref{fig:steps}c-d and its zoom is shown in Figure \ref{fig:steps}e-f.

\begin{figure}
     \centering
     \subfloat[][curve for 0, 1, and 2 magnets]{\includegraphics[width=0.45\textwidth]{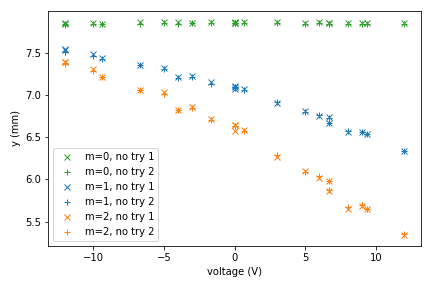}\label{fig:staticThreeAll}}
     \subfloat[][zoom of curve without magnet]{\includegraphics[width=0.45\textwidth]{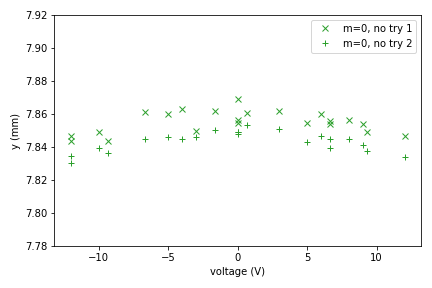}\label{fig:staticThreeZoom}}
     \caption{The static characteristics for the actuator with zero, one, and two magnets.}
     \label{fig:staticThree}
\end{figure}

The steady-state characteristics were calculated by averaging the output from the last \SI{0.25}{\second} from the single-step response and it is visible in Figure \ref{fig:staticThreeAll} for none, one, and two permanent magnets. The results suggest that the range of movement is increased by a number of permanent magnets. The lack of a magnet gives a very small range of movement which is shown in Figure \ref{fig:staticThreeZoom}. The exact range of static movement is as follows: none - \SI{0.025}{\milli\meter}, one - \SI{1.185}{\milli\meter}, and two - \SI{2.040}{\milli\meter}. Furthermore, the number of permanent magnets provides a different initial offset for zero voltage. The permanent magnet causes the up and down movement of the membrane for positive/negative voltage. In the case of a standalone membrane (without a magnet), the movement is unidirectional. This effect is caused by different forces which cause the movement. In the case of the addition of a permanent magnet, the force between the magnet, membrane, and electromagnet is present. In the case without a permanent magnet, only the reluctance effect cause the movement.

\subsection{Identification Results}

\begin{figure}
     \centering
     \subfloat[][global coordinate $y$]{\includegraphics[width=0.45\textwidth]{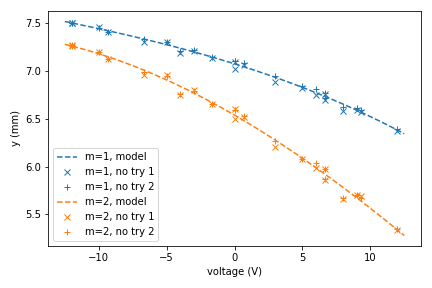}\label{fig:staticLocal}}
     \subfloat[][local coordinate $\Delta y$]{\includegraphics[width=0.45\textwidth]{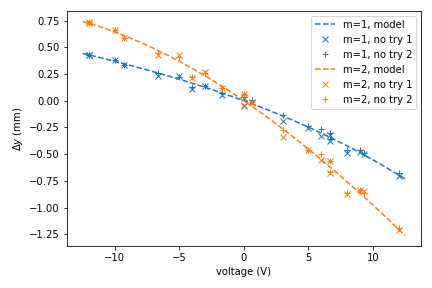}\label{fig:staticGlobal}}
     \caption{The static responses of the actuator in global and local coordinates.}
     \label{fig:static}
\end{figure}

The identification is performed only for one and two permanent magnets. The algorithm was computed in Python \cite{2020SciPy-NMeth}. Firstly the static nonlinearity is calculated for the characteristics presented in Figure \ref{fig:staticThreeAll}. The obtained coefficients are given in Table \ref{tab:polynomialParameters}. The coefficient $p_0$ is equal to the working point $y_0$ (for $u$=0). It is visible that an increase in the magnet's number causes a decrease in the initial space between MRE membrane and the electromagnet. The visualization of results is presented in Figure \ref{fig:static} where it is visible that static characteristic is not fully symmetric for positive and negative voltage. 

\begin{table}
 \caption{The parameters of polynomial representing the static nonlinearity.}
  \centering
  \begin{tabular}{ccccc}
    \toprule
    & & \multicolumn{3}{c}{Polynomial coefficient}                   \\
    \cmidrule(r){2-5}
    Case     & $p_3$     & $p_2$ & $p_1$ & $p_0$ \\
    \midrule
    Single Magnet & \num{-1.709d-5} & \num{-9.229d-4} & \num{-4.435d-2} & \num{7.077} \\
    Double Magnet & \num{1.450d-5} & \num{-1.663d-3} & \num{-8.240d-2} & \num{6.539} \\
    \bottomrule
  \end{tabular}
  \label{tab:polynomialParameters}
\end{table}

\begin{table}
 \caption{The parameters of transfer function based on identification procedure.}
  \centering
  \begin{tabular}{cccc}
    \toprule
    & & \multicolumn{2}{c}{Median}                   \\
    \cmidrule(r){3-4}
    Parameter     & Trans. Func.     & Single Magnet & Doubled Magnet \\
    \midrule
    gain $\tilde{k}$                    & fast & 0.76 & 0.77 \\
    decay rate $\tilde{\alpha}$         & fast & 39.98 & 32.91 \\
    angular frequency $\tilde{\omega}$  & fast & 246.11 & 218.98 \\
    pole $\tilde{s}_0$                  & slow & 0.13 & 0.14 \\
    \bottomrule
  \end{tabular}
  \label{tab:transferFunctions}
\end{table}

\begin{figure}
     \centering
     \subfloat[][single magnet, step j=6, $u^j_{\text{pre}}$=7.8, $u^j_{\text{post}}=-4.0(V)$]{\includegraphics[width=0.45\textwidth]{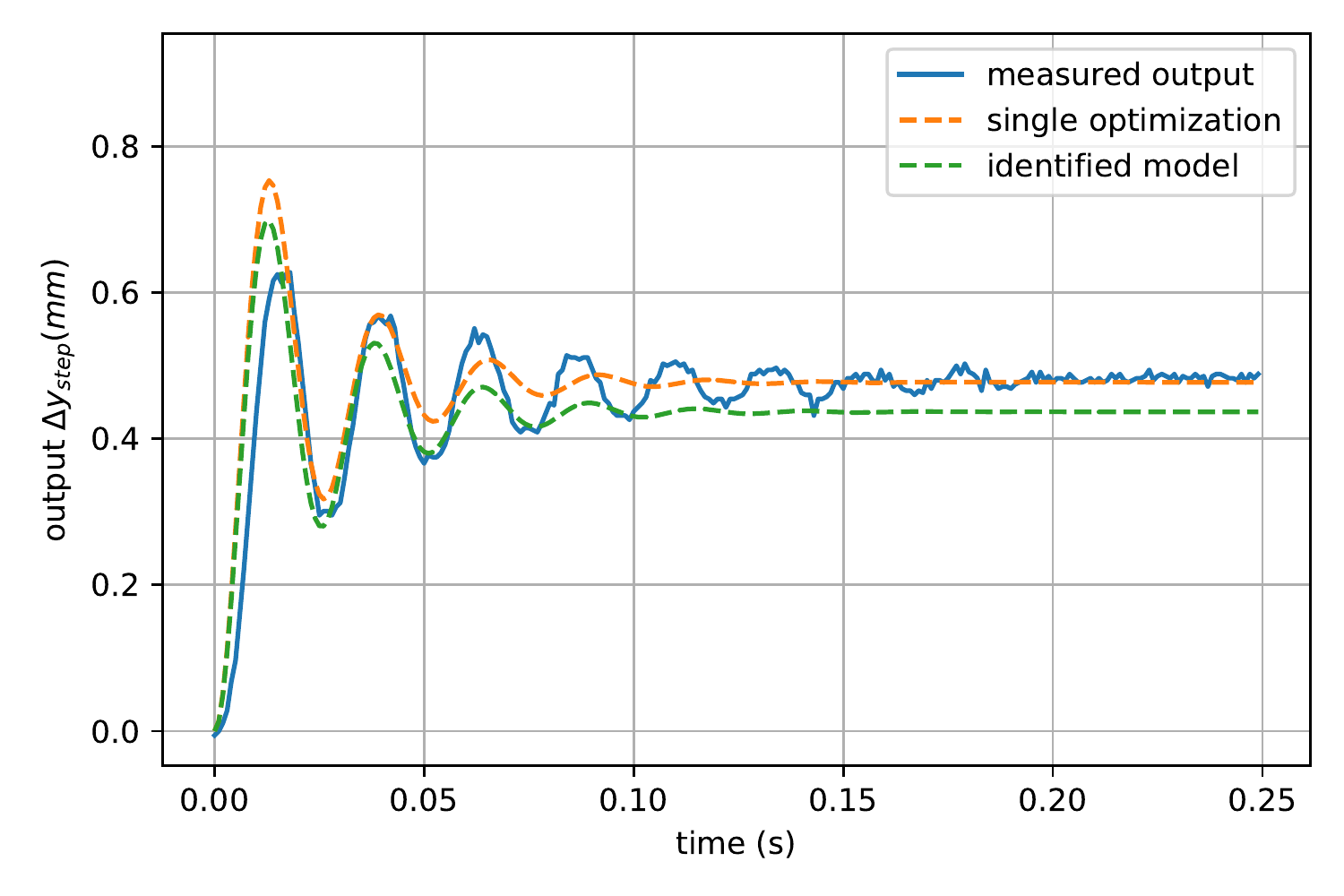}}
     \subfloat[][double magnet, step j=6, $u^j_{\text{pre}}$=7.8, $u^j_{\text{post}}=-4.0(V)$]{\includegraphics[width=0.45\textwidth]{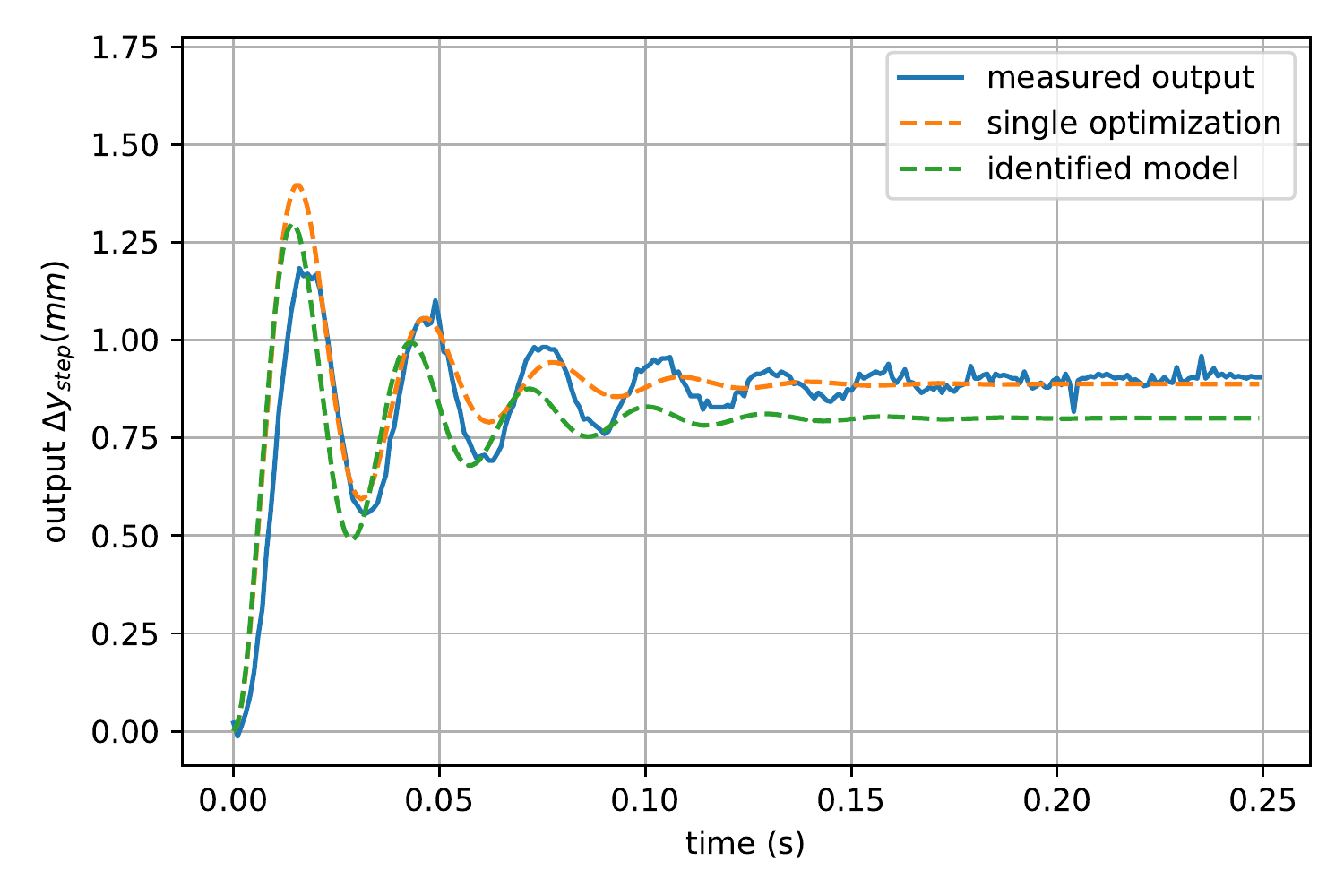}} \\
     \subfloat[][single magnet, step j=14, $u^j_{\text{pre}}=$-9.9, $u^j_{\text{post}}=-5.0(V)$]{\includegraphics[width=0.45\textwidth]{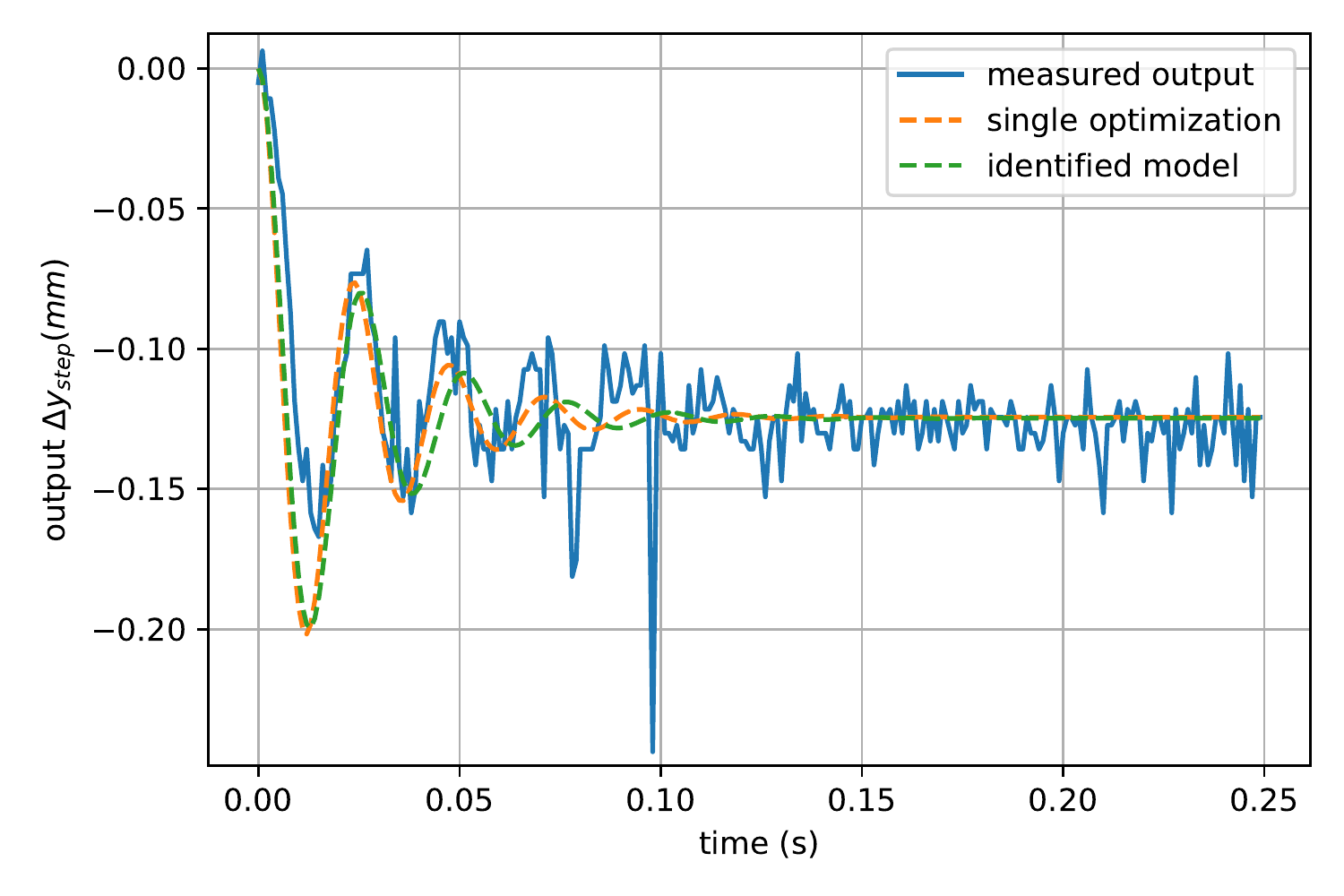}}
     \subfloat[][double magnet, step j=14, $u^j_{\text{pre}}=$-9.9, $u^j_{\text{post}}=-5.0(V)$]{\includegraphics[width=0.45\textwidth]{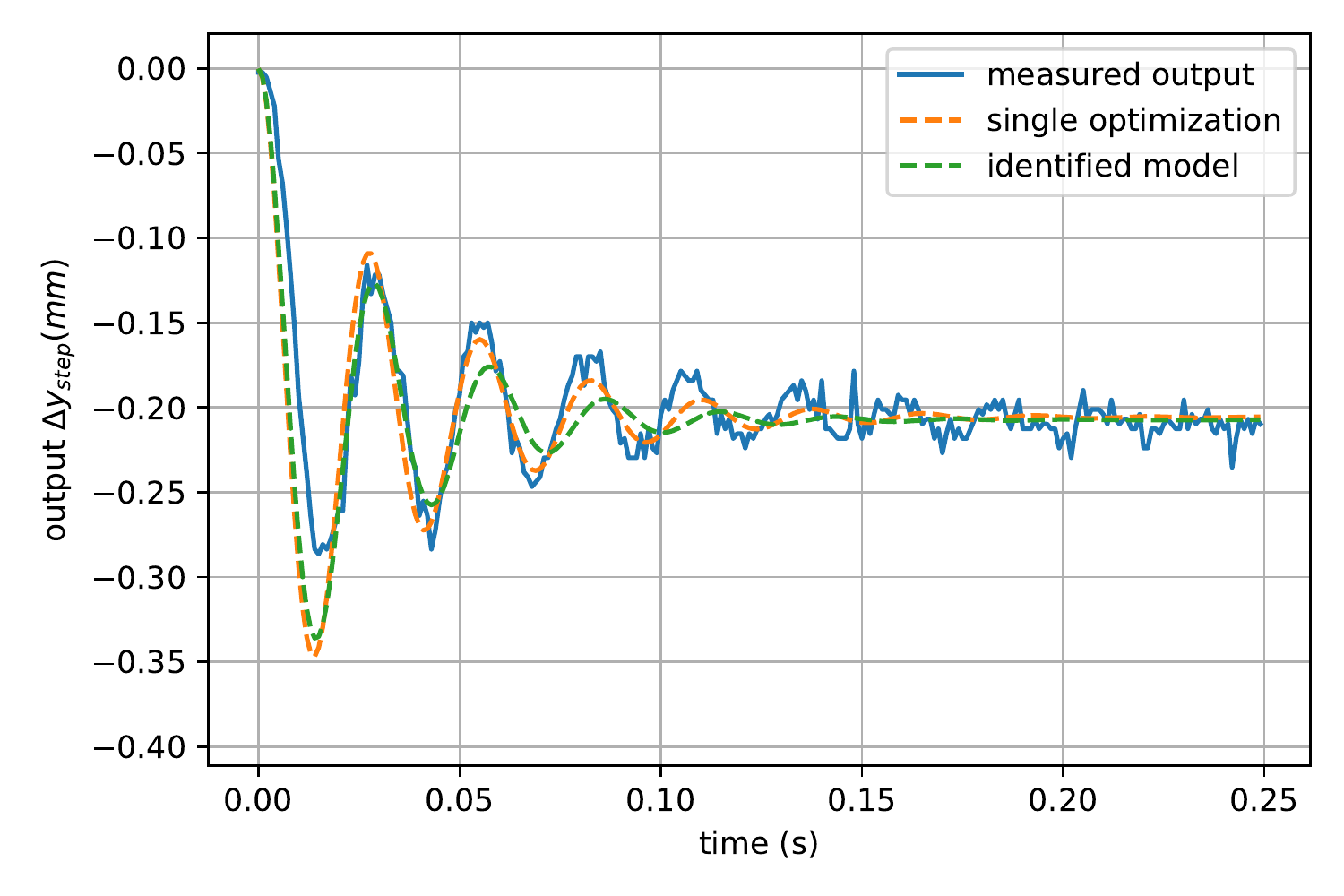}}
     \caption{The responses obtained by optimization for transfer function $G_{fast}(s)$.}
     \label{fig:compareGfast}
\end{figure}

\begin{figure}
     \centering
     \subfloat[][single magnet, step j=6, $u^j_{\text{pre}}=$7.8, $u^j_{\text{post}}=-4.0(V)$]{\includegraphics[width=0.45\textwidth]{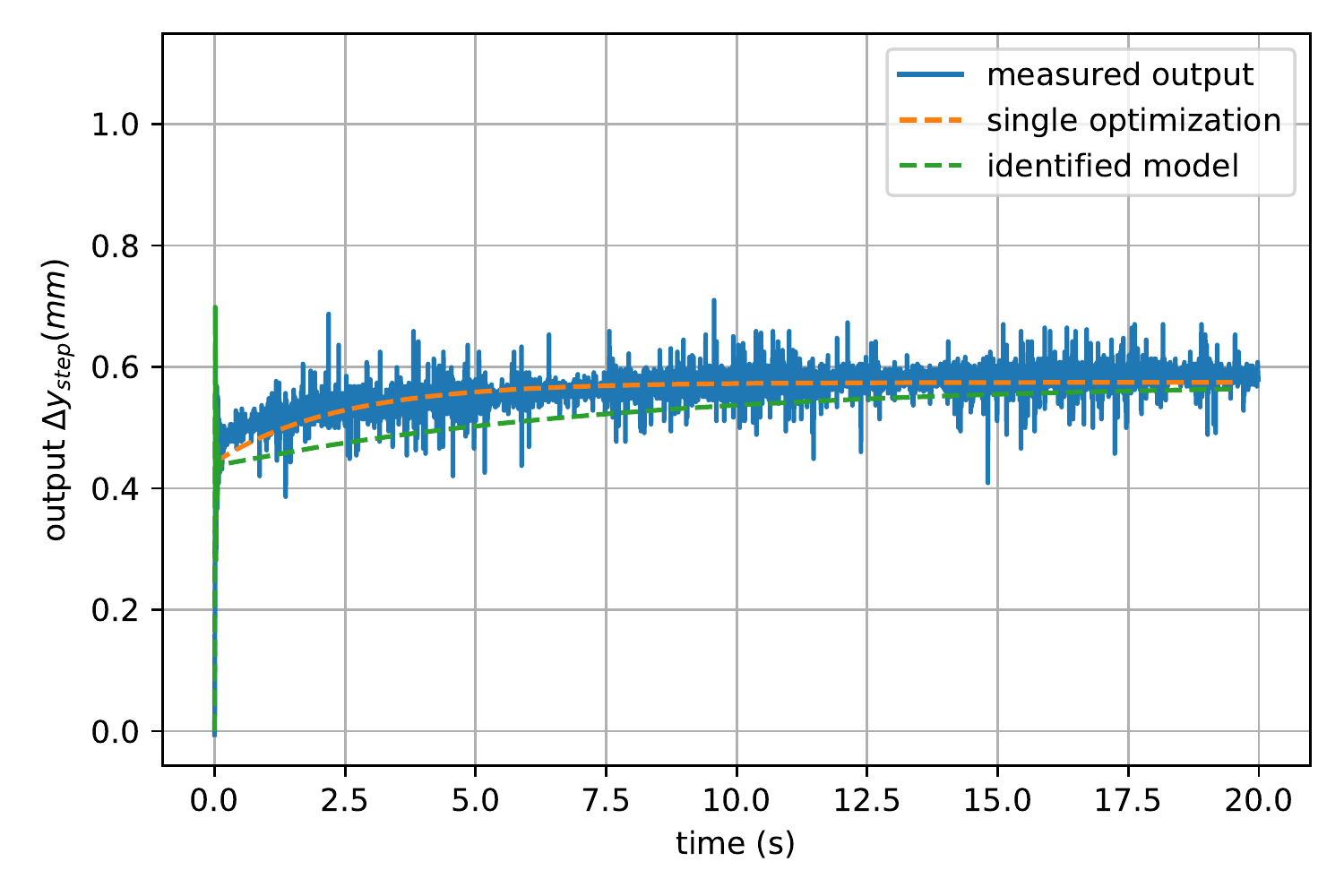}}
     \subfloat[][double magnet, step j=6, $u^j_{\text{pre}}=$7.8, $u^j_{\text{post}}=-4.0(V)$]{\includegraphics[width=0.45\textwidth]{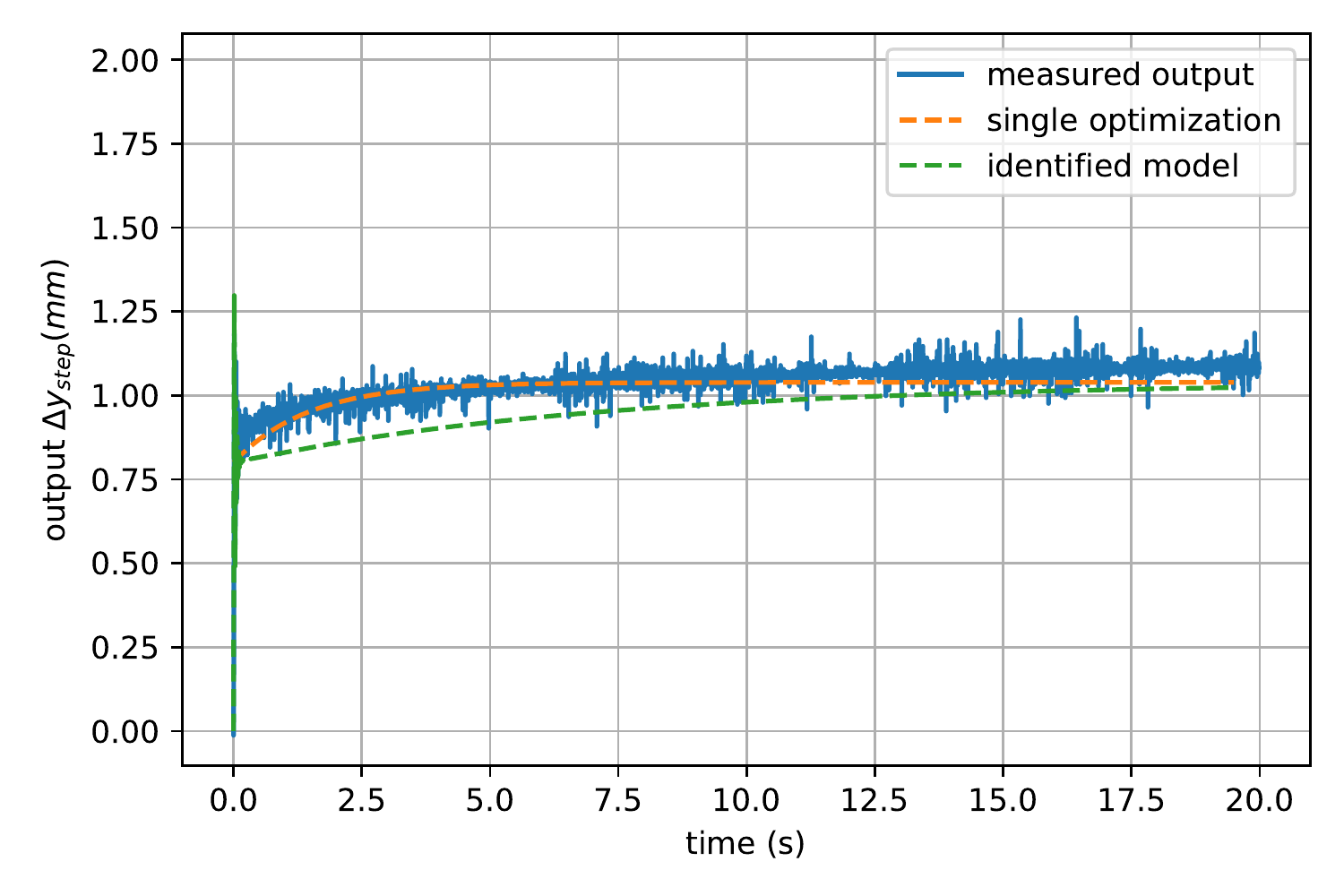}} \\
     \subfloat[][single magnet, step j=14, $u^j_{\text{pre}}=$-9.9, $u^j_{\text{post}}=-5.0(V)$]{\includegraphics[width=0.45\textwidth]{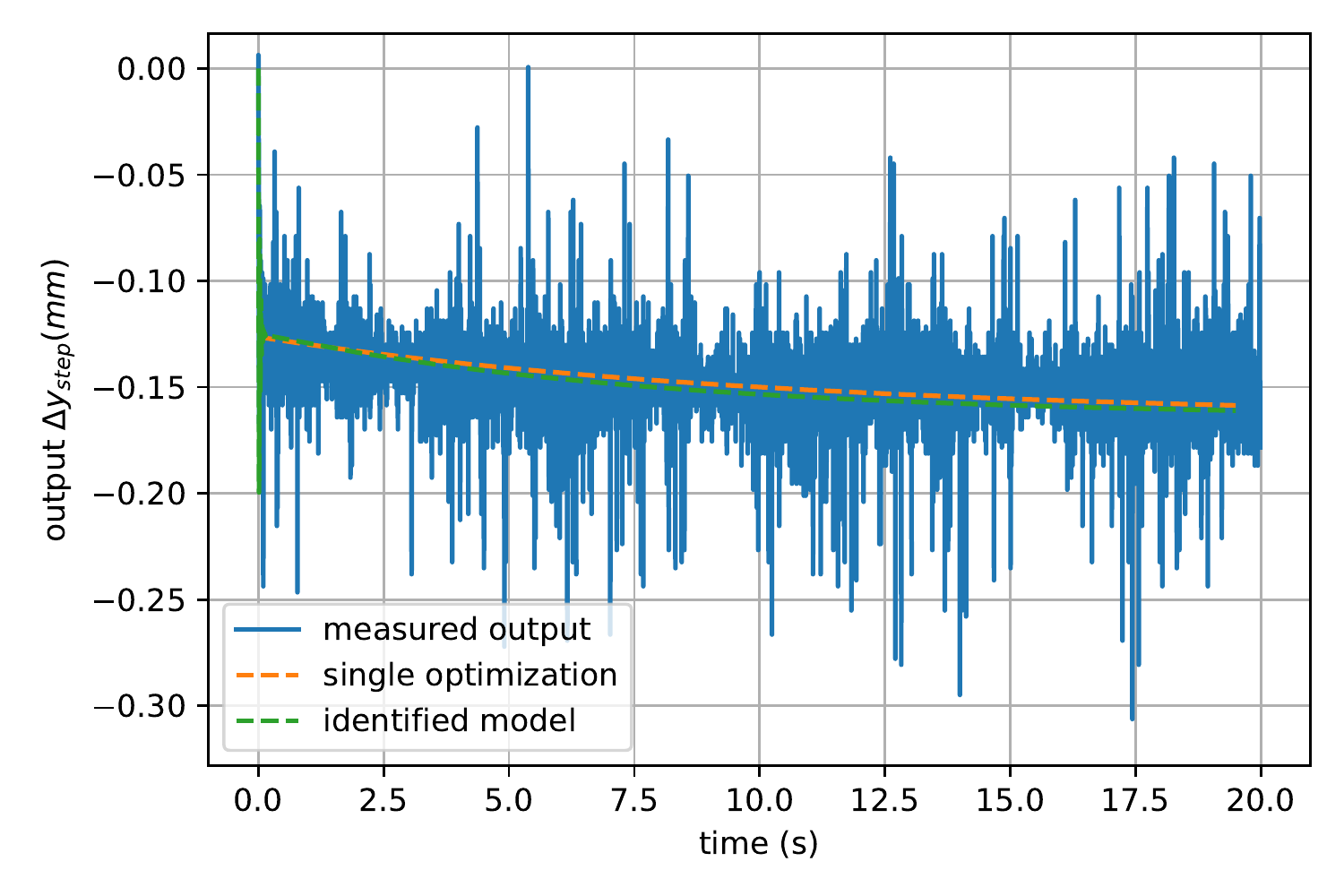}}
     \subfloat[][double magnet, step j=14, $u^j_{\text{pre}}=$-9.9, $u^j_{\text{post}}=-5.0(V)$]{\includegraphics[width=0.45\textwidth]{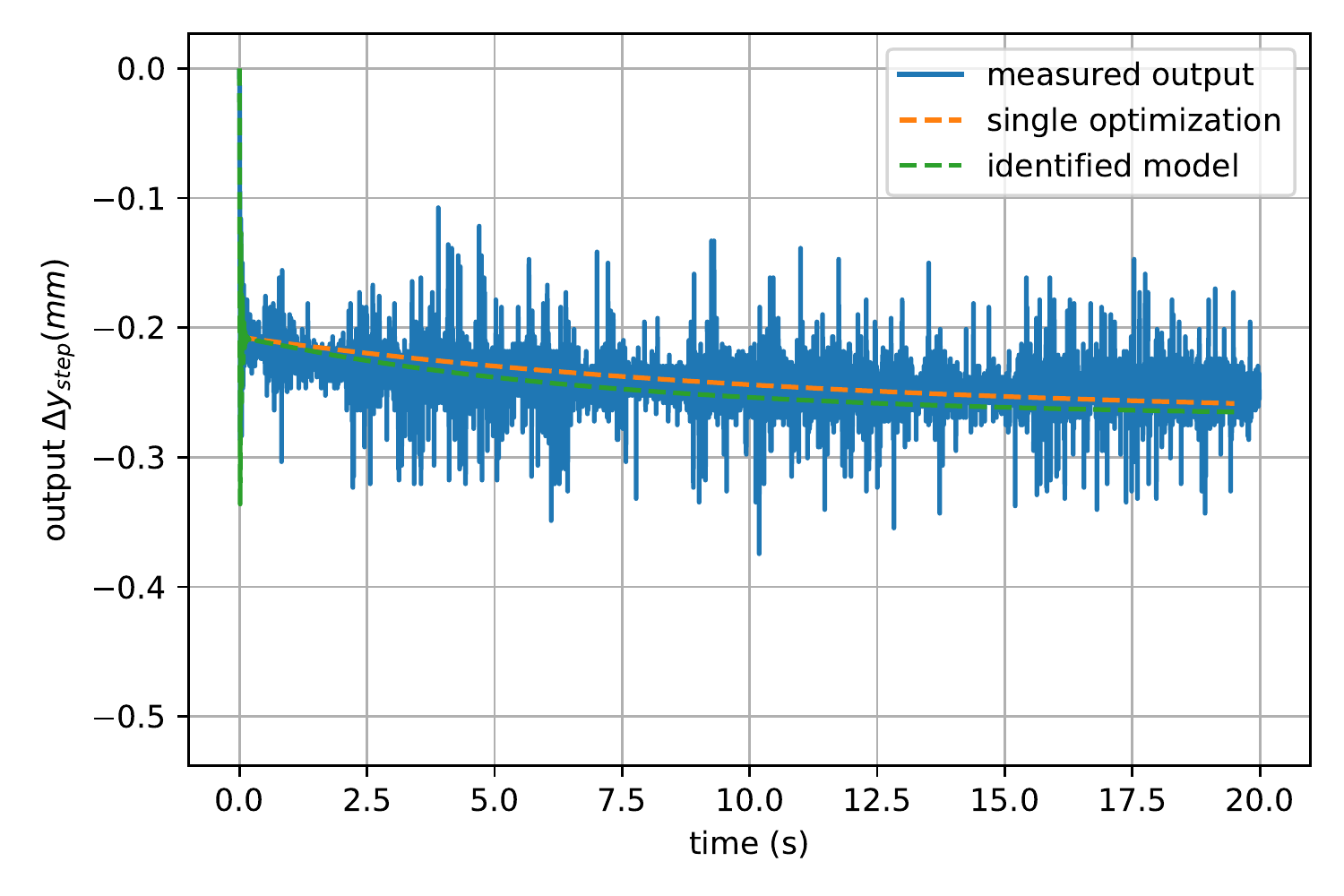}}
     \caption{The responses obtained by optimization for transfer function $G_{slow}(s)$.}
     \label{fig:compareGslow}
\end{figure}

\begin{figure}
     \centering
     \subfloat[][single magnet, fast-slow transfer function]{\includegraphics[width=0.45\textwidth]{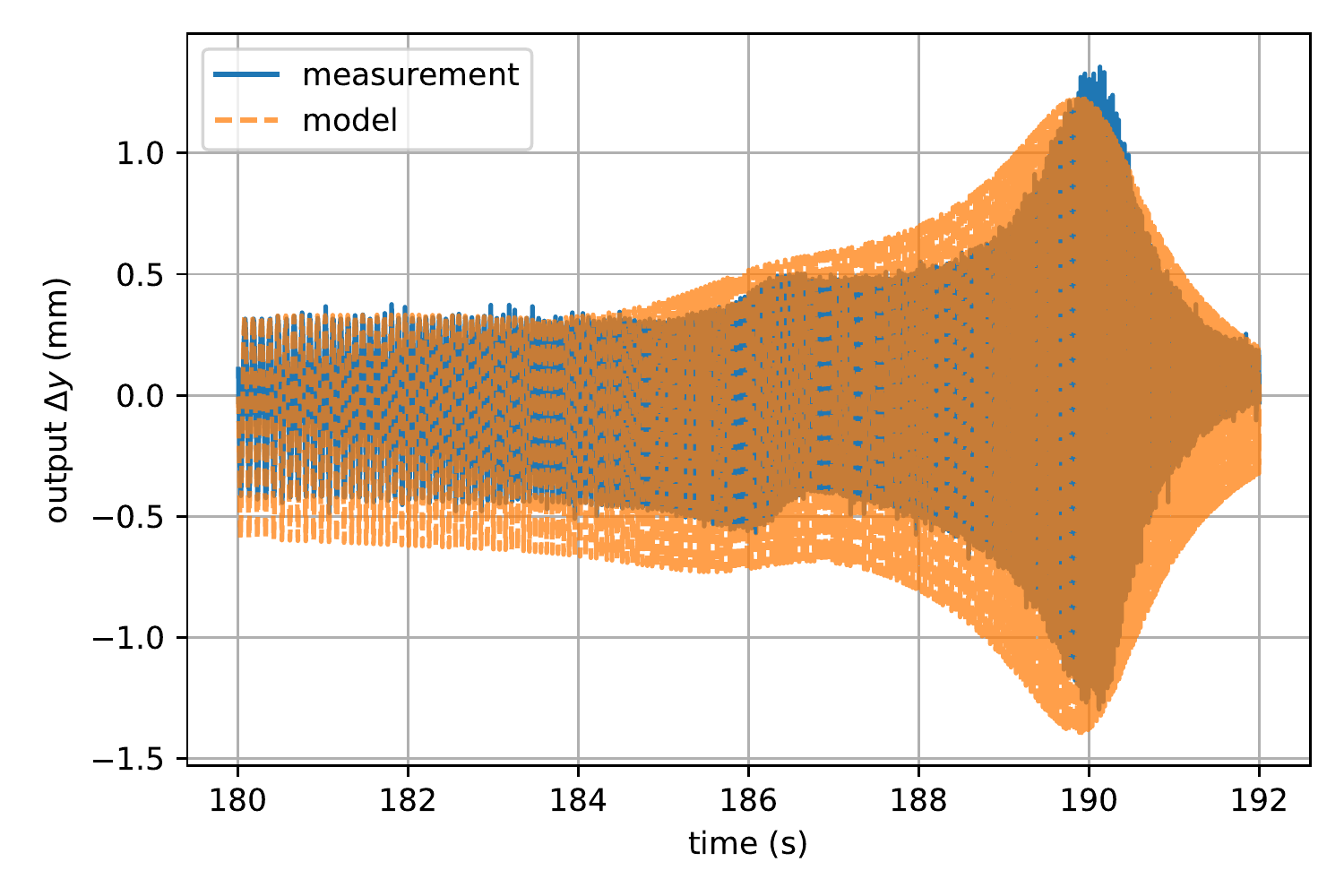}}
     \subfloat[][double magnet, fast-slow transfer function]{\includegraphics[width=0.45\textwidth]{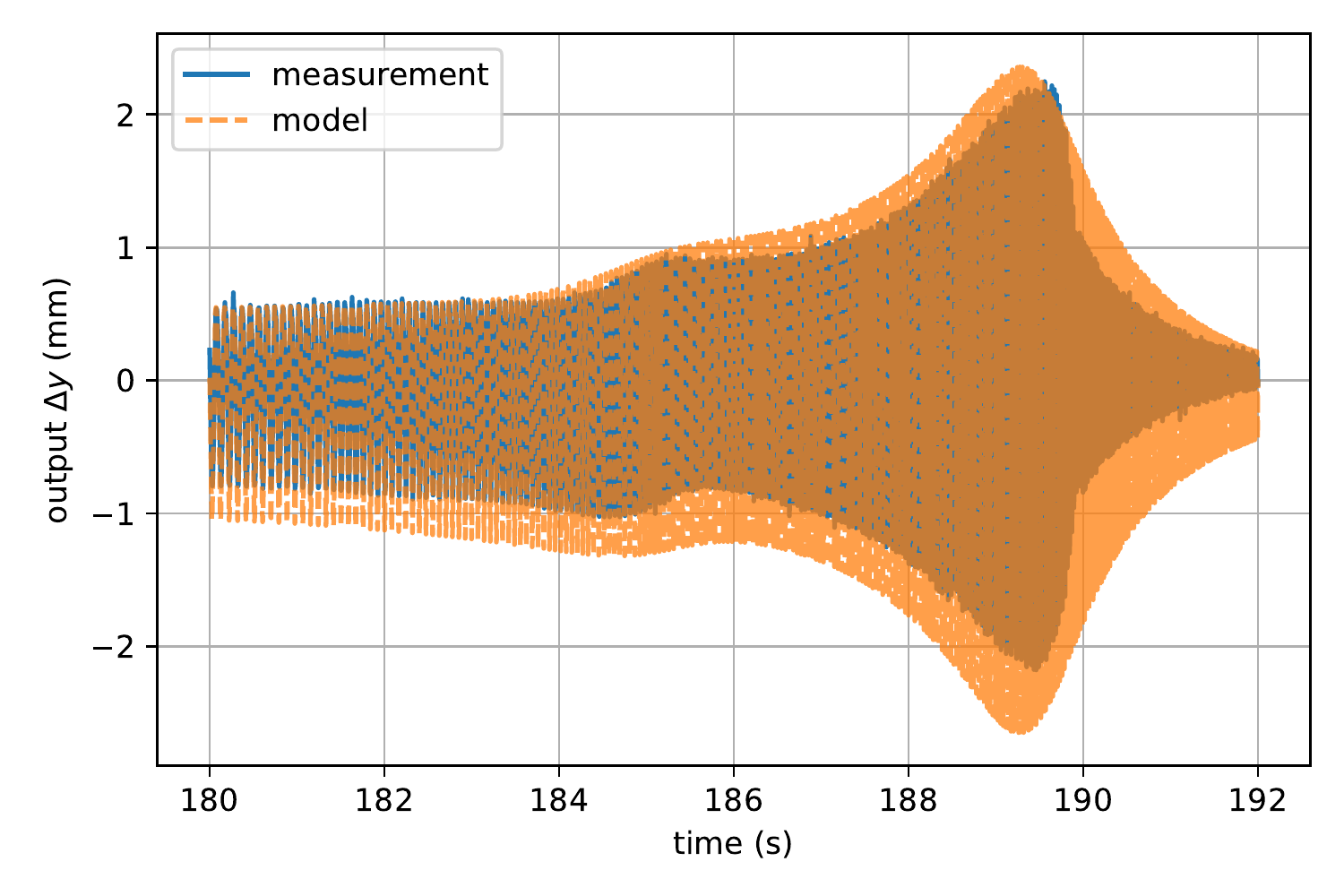}}\\
     \subfloat[][single magnet, single transfer function]{\includegraphics[width=0.45\textwidth]{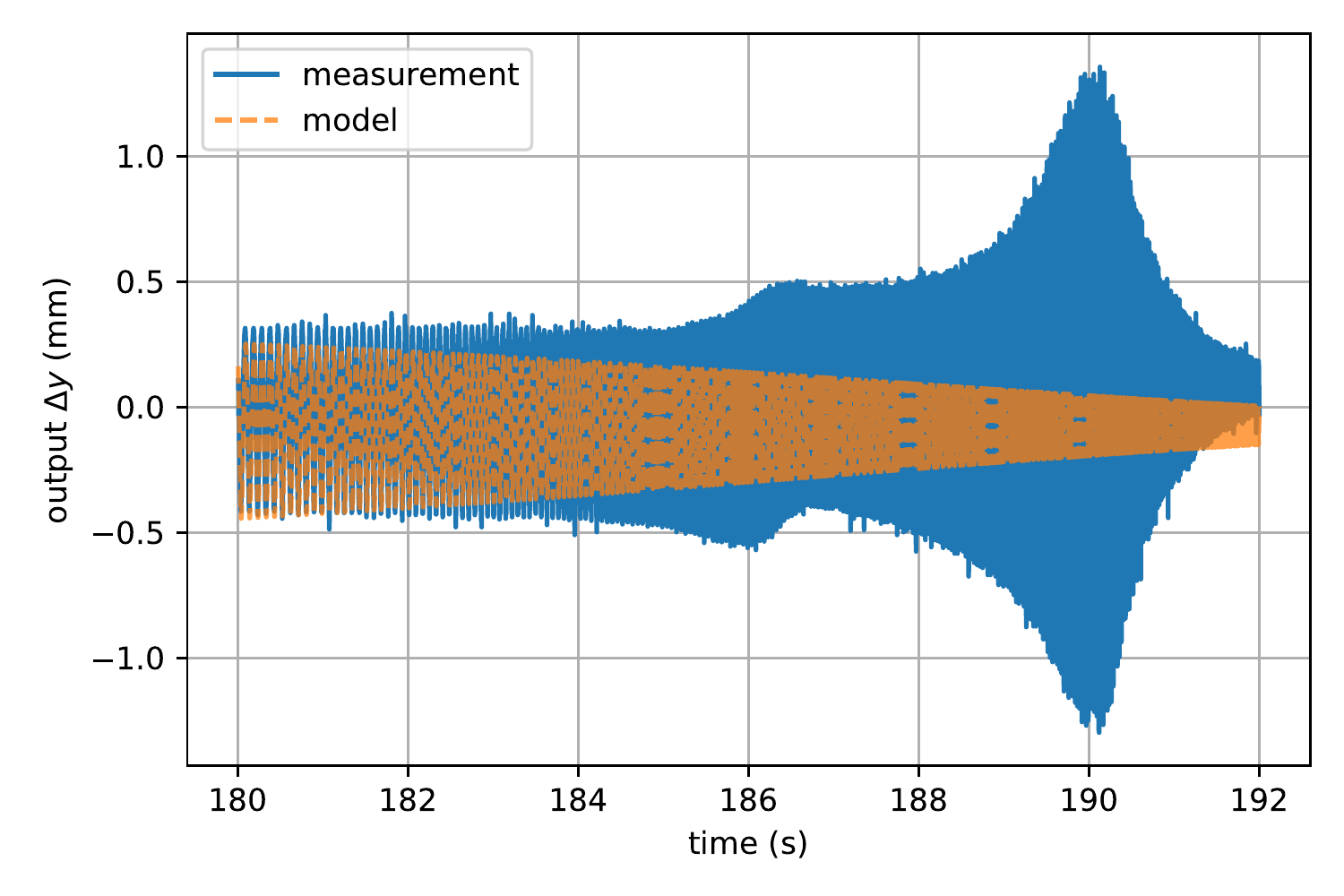}}
     \subfloat[][double magnet, single transfer function]{\includegraphics[width=0.45\textwidth]{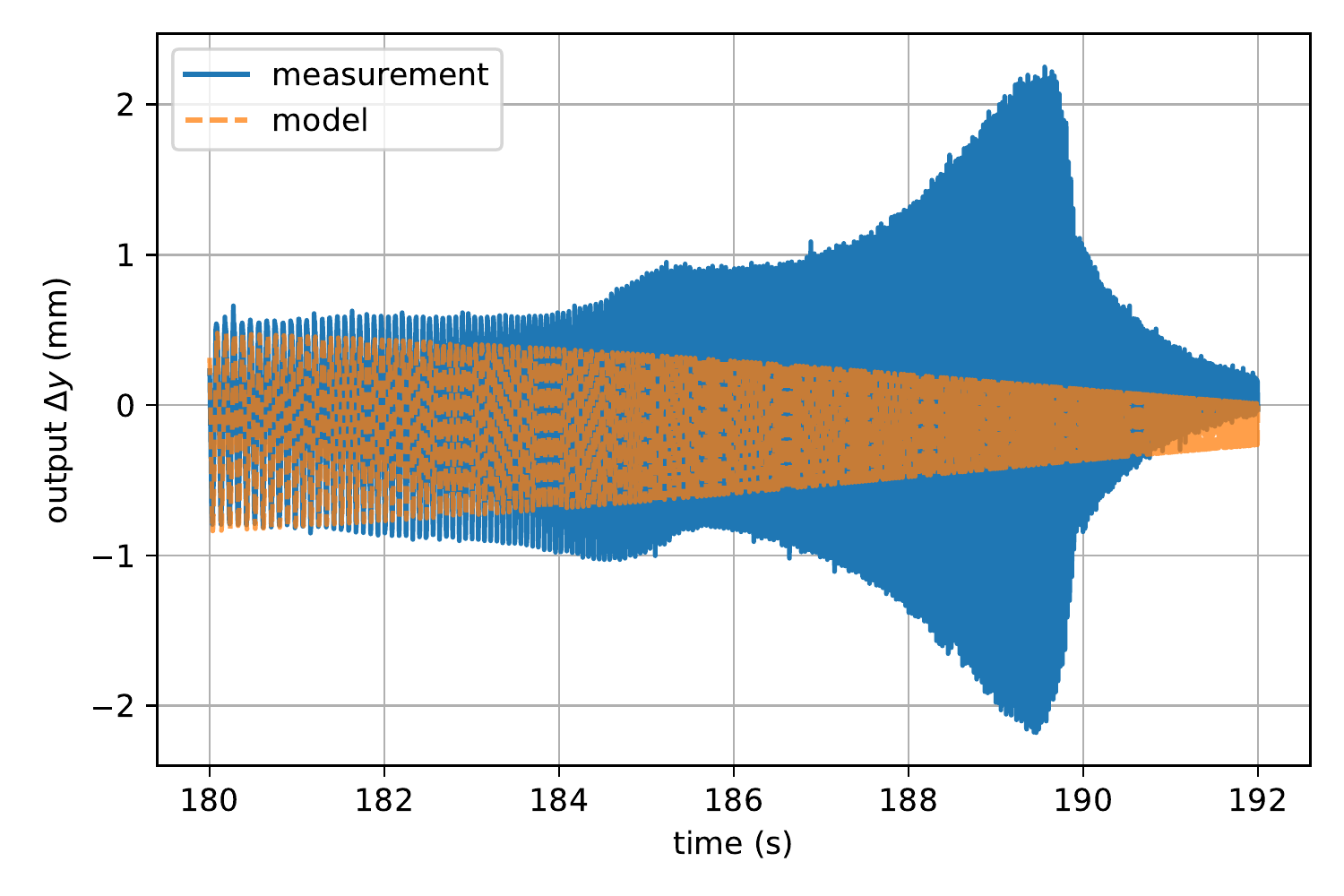}}
     \caption{The validation of both models for chirp signal. The identification method with slow and fast transfer function (a-b) and the single transfer function method (c-d). }
     \label{fig:validationChirp}
\end{figure}

The transfer function is identified by performing the procedure presented in Section \ref{se:hammersteinModelAndIdentification}. The parameters of transfer function are given in Table \ref{tab:transferFunctions}. The results of identification are visible in Figures \ref{fig:compareGfast} and \ref{fig:compareGslow}. The identified model well describes the experimental data. It covers the oscillations and decay rate. Furthermore, to check the model on non-learning data, the validation signal was applied to the actuator and the response was measured. The comparison is illustrated in Figure \ref{fig:validationChirp}a-b which shows that the model can predict well the resonance.

In Figure \ref{fig:validationChirp}c-d the comparison to the standard method with a single transfer function is shown. The identification is performed for a transfer function:
\begin{equation}
G(s) = \frac{b_1s + b_0}{s^3 + a_2 s^2 + a_1 s + a_0}. 
\end{equation}
The identification was run for a single and double magnet case. Because of a single transfer function, the dynamic data was not split into two cases. The identification process is run by the instrumental variable for continuous systems based on the implementation in the Identification Toolbox in Matlab. The process finished with the success of obtaining the fit to estimation data above \SI{92}{\percent}. However, it is visible that the long time of relaxation process dominates the identification and hence the resonance is not identified correctly. These results suggest that the identification with two steps is more suitable for the presented type of problem in this work. 

Observations of the model parameters presented in Table \ref{tab:parameters} imply that the decay rate and resonant frequency decreased with the number of magnets. It could be explained by increasing the stiffness with the application of a higher magnetic field from the permanent magnet. In general, it is known that the storage and loss modulus which represents the mechanical properties of the membrane are increasing with the magnetic field \cite{Bose:2021}. On the other hand, the double magnet construction has a significantly greater moving range which shows that flux from permanent magnet has a significant role in the force production.

\begin{figure}
     \centering
     \subfloat[][single magnet]{\includegraphics[width=0.45\textwidth]{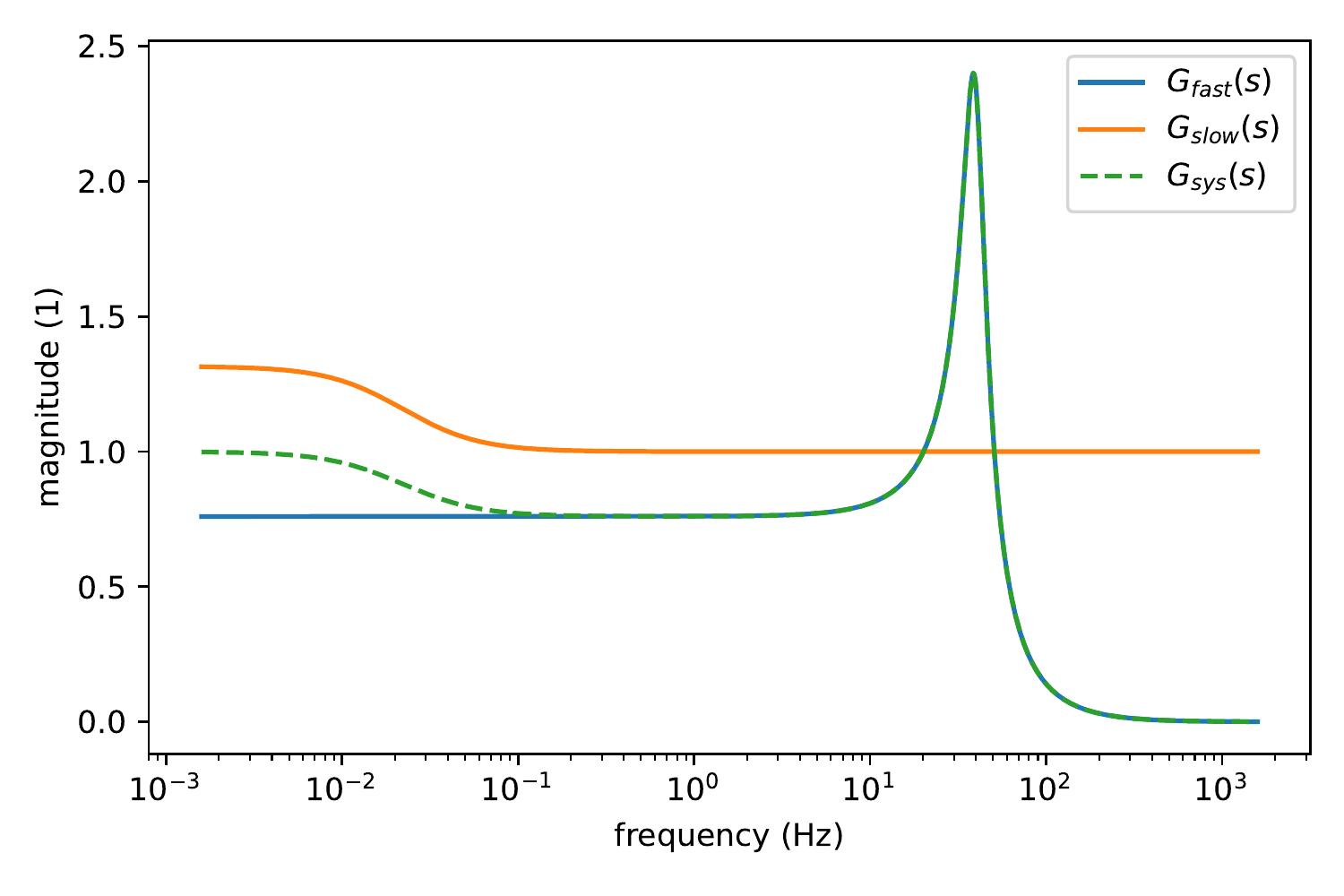}}
     \subfloat[][double magnet]{\includegraphics[width=0.45\textwidth]{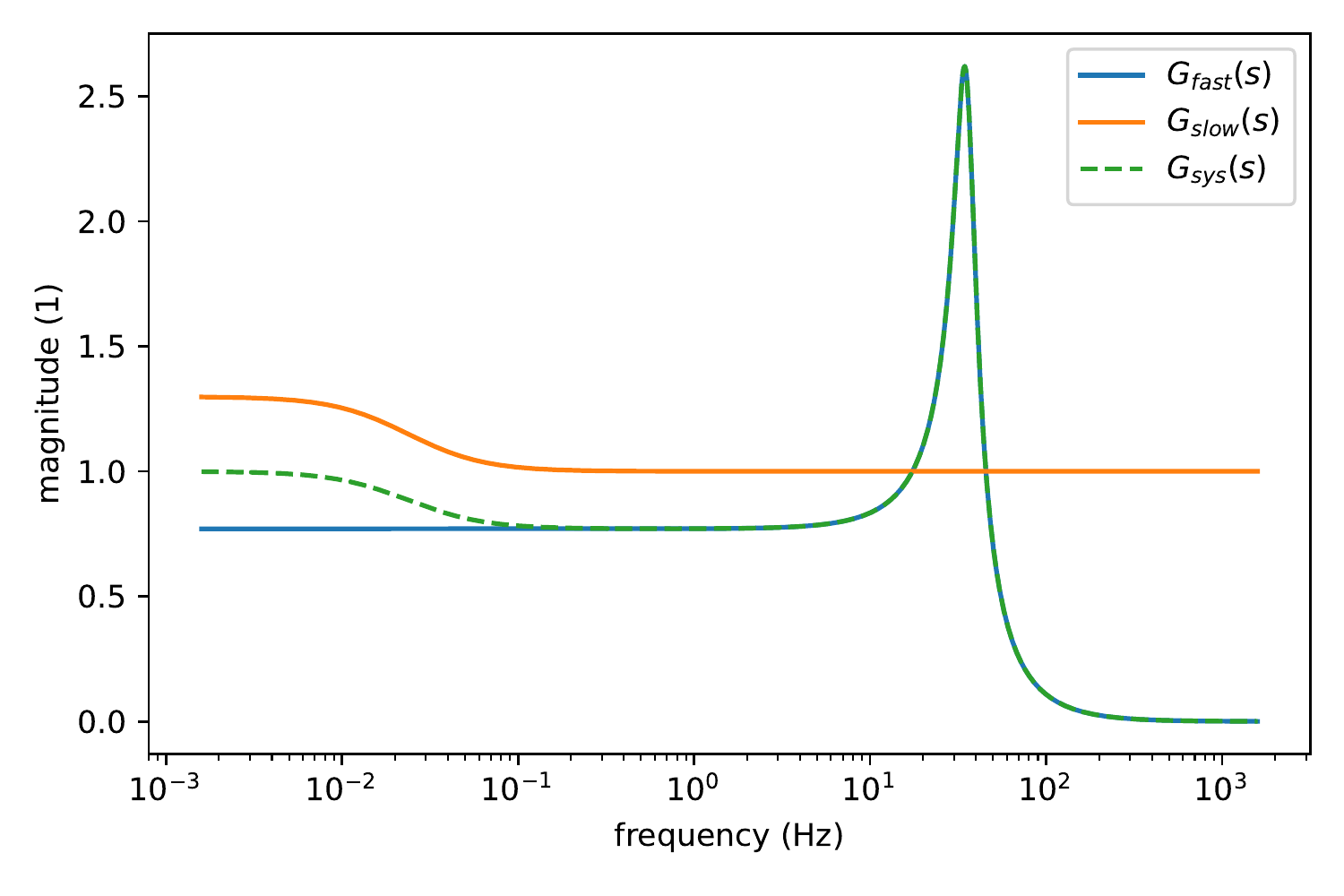}}
     \caption{The bode plots of identified plants.}
     \label{fig:bode}
\end{figure}

The bode plots of the transfer function are presented in Figure \ref{fig:bode}. It is visible that the system satisfies the assumptions given in the identification procedure. This means that $G_{slow}(s)$ is almost 1 for high frequencies and $G_{fast}(s)$ is almost $k$ for low frequencies.

\subsection{Genaral Properties of Actuator and Its Further Applications}

In general, the proposed actuator has similar responses to DEAP actuator presented in work \cite{Rizzello:6867294,Bernat:9293025}. It also is axisymmetrical. However, the main difference is in the actuation principle - magnetic field in the presented work versus electric field for DEAP actuators. The response of both actuators has similar features like oscillations and relaxation time caused by silicone properties. The potential applications are the pump systems or loudspeakers, similar to DEAP actuators \cite{Bar-Cohen:01,SIDERIS2020111915}. Another possibility is the varying-stiffness push-button known from intelligent transducers \cite{Kim:2007}.

\section{Conclusion}
\label{se:conclusion}

In this work, the novel concept of a magnetorheological actuator is presented. The application of permanent magnets increased significantly the moving range of device. The axisymmetrical concept of the actuator gives a wide range of possible applications. The model has two-time scale responses which are difficult to identify by standard procedures. By the application of separate identification of fast and slow dynamics, it is possible to more accurately describe the model. 

\section*{Acknowledgments}
This research was funded by Ministry of Education and Science, grant number 0211/SIGR/6434

\bibliographystyle{unsrt}  
\bibliography{references}

\end{document}